\documentclass[10pt,letterpaper]{article}
\usepackage[top=0.85in,left=2.75in,footskip=0.75in]{geometry}

\usepackage{changepage}

\usepackage[utf8]{inputenc}

\usepackage{textcomp,marvosym}

\usepackage{fixltx2e}

\usepackage{amsmath,amssymb}

\usepackage{cite}

\usepackage{nameref}%,hyperref}

\usepackage{microtype}
\DisableLigatures[f]{encoding = *, family = * }

\usepackage{rotating}

\usepackage{pdfpages}

\raggedright
\usepackage[aboveskip=1pt,labelfont=bf,labelsep=period,justification=raggedright,singlelinecheck=off]{caption}

\makeatletter
\renewcommand{\@biblabel}[1]{\quad#1.}
\makeatother

\date{}

\usepackage{lastpage,fancyhdr,graphicx}
\usepackage{epstopdf}
\fancyheadoffset[L]{2.25in}
\fancyfootoffset[L]{2.25in}
\begin{document}
\vspace*{0.35in}

% Please capitalize all terms in the title except conjunctions, prepositions, and articles.
\begin{flushleft}
{\Large
\textbf\newline{Quantifying the Consistency of Scientific Databases}
}
\newline
\\
Lovro Šubelj\textsuperscript{1,*},
Marko Bajec\textsuperscript{1},
Biljana Mileva Boshkoska\textsuperscript{2},
Andrej Kastrin\textsuperscript{2},
Zoran Levnajić\textsuperscript{1,2}
\\
\bigskip
\bf{1} Faculty of Computer and Information Science, University of Ljubljana, Ljubljana, Slovenia
\\
\bf{2} Faculty of Information Studies in Novo mesto, Novo mesto, Slovenia
\\
\bigskip

% Insert additional author notes using the symbols described below. Insert symbol callouts after author names as necessary.
% 
% Remove or comment out the author notes below if they aren't used.
%
% Primary Equal Contribution Note
%\Yinyang These authors contributed equally to this work.

% Additional Equal Contribution Note
% Also use this double-dagger symbol for special authorship notes, such as senior authorship.
%\ddag These authors also contributed equally to this work.

%\textcurrency a Insert current address of first author with an address update
% \textcurrency b Insert current address of second author with an address update
% \textcurrency c Insert current address of third author with an address update


% Group/Consortium Author Note
%\textpilcrow Membership list can be found in the Acknowledgments section.

% Use the asterisk to denote corresponding authorship and provide email address in note below.
* lovro.subelj@fri.uni-lj.si
\end{flushleft}

\section*{Abstract}
Science is a social process with far-reaching impact on our modern society. In the recent years, for the first time we are able to scientifically study the science itself. This is enabled by massive amounts of data on scientific publications that is increasingly becoming available. The data is contained in several databases such as Web of Science or PubMed, maintained by various public and private entities. Unfortunately, these databases are not always consistent, which considerably hinders this study. Relying on the powerful framework of complex networks, we conduct a systematic analysis of the consistency among six major scientific databases. We found that identifying a single ``best'' database is far from easy. Nevertheless, our results indicate appreciable differences in mutual consistency of different databases, which we interpret as recipes for future bibliometric studies.

%\linenumbers

\section*{Introduction}

Science is a human endeavor. As such, it benefits from all virtues and suffers from all paradoxes inherent to humans. Among these are the old problems of appreciating and measuring research achievements~\cite{DRW2014}. When judging what is and what is not scientifically interesting or important, scientists are not just subjective, but often offer arguments that stem from poor understanding of the academic culture and tradition in fields other than their own. In the age of Big data, \textit{science of science} is emerging as an attempt to scientifically examine the science itself~\cite{GUSA2005,PF2014}. This young field has potential to answer some of the oldest questions about scientific progress, such as elucidating the sociological mechanisms leading to new discoveries~\cite{UMSJ2013,WJU2007,MILOJEVIC2014}, or establishing a platform for objectively quantifying scientific impact~\cite{PFP2014,PF2014,YWZFD2014}. These insights are also useful in building realistic scenarios of future development of science and its impact on our lives~\cite{WJU2007,PFP2014,SB2014}. Science of science also receives attention from policy makers~\cite{sosp}. Indeed, being able to fairly evaluate and compare scientific outputs enables the community to improve the funding strategies and target them towards achievable goals. It also provides a framework to quantify the research impact resulting from a given investment~\cite{SB2014}.

The dynamics of science is articulated through a constant influx of scientific publications, primarily research papers. Appearing in a variety of journals, papers are interrelated in intricate ways, governed by complex patterns of co-authorships (collaborations)~\cite{New01e} and citations~\cite{SB13}. Hidden in these patterns are the answers to many pondering questions: Which papers set the new trends~\cite{KYC12a}? Can their eventual impact be recognized early upon publication~\cite{WSB13}? How does interdisciplinary research arise and what are the best ways to stimulate it~\cite{SA12a}? Extracting these answers calls for new methodologies of untangling these complex patterns from scientific databases such as Web of Science or arXiv. The only way to  exploit the rapid growth of bibliometric (scientometric) data, is to parallel it with equally rapid growth and improvement of methodologies aimed at efficiently mining them.

In this context, the framework of networks (graphs) has been recognized as an elegant tool for representing and analyzing complex systems~\cite{BBV2008,New10}. In a variety of fields ranging from computer science and physics to sociology and biology, this approach has provided paradigm-shifting results~\cite{EK2010,SFB11}. In particular, scientific databases can be represented as complex networks by identifying publications or authors as network nodes and modeling their bibliometric relationships as network links~\cite{New01e,Per10a}. Relying on this paradigm, intense research efforts over the last decade provided novel quantitative findings on dynamics and evolution of science. Besides being suited for analyzing the emergence of interdisciplinarity~\cite{LLPP14}, this framework gave insights into new ways of estimating scientific impact~\cite{WSB13,KFMWH11}, opened a window into the communities among scientists~\cite{ELP11,PS12a}, or enabled novel approaches to study the evolution of science~\cite{EF11,AHL12}.

However, despite promising results and increasing availability of data, the core obstacle is the lack of a universal scientific database with all data systematically stored. Instead, there are several databases, each relying on its own practice in storing, organizing and tracking bibliometric data, including Web of Science, arXiv, PubMed etc. Moreover, none of the datasets is free from errors, mostly occurring due to different referencing styles or typos in authors names (in particular names utilizing non-English characters), which often lead to incorrectly recorded collaborations and citations. This in practice means that each bibliometric study in itself unavoidably carries some degree of bias, resulting from the choice of the database. On top of this comes the fact that different fields usually have different collaboration and citation cultures, which further complicates issue of objectively comparing different scientific fields.

On the other hand, researchers is bibliometrics usually work relying on the database at their disposal. Finding additional data is often difficult and sometimes expensive. While the construction of a universal database is an ambitious goal, we recognize that the bibliometric community will benefit from a critical comparison of the available databases. Of course, since there is no ``ground truth'' to tell between the reliable and non-reliable databases, the best we can do is to systematically examine and quantify the consistency among different scientific databases. We here conduct a detailed analysis of the consistency among six major scientific databases, employing three different paradigms (categories) of bibliometric networks (paper citation, author citation and collaboration). This amounts to a major methodological and empirical extension of our earlier paper~\cite{SFB14}: additional datasets and network paradigms are considered, and findings confirmed by complementary analyses. Our results consist of an approximate quantification of consistency between the six databases that hold within each network category. Our study aims at being helpful to colleagues when choosing the most suitable network paradigm.

% Results and Discussion can be combined.
\section*{Results}

We obtained the data on co-authorships and citations from the following six databases: American Physical Society (APS), Web of Science (WoS), DBLP, PubMed, Cora and arXiv. Since some databases are very large (e.g. WoS), we were unable to include them entirely. Nevertheless, we made sure that the dataset from each database is representative of it in terms of papers and citations (see Methods). From each database we constructed three bibliometric networks using the following three network paradigms (categories):
\begin{itemize}
	\item P$\rightarrow$P, directed paper citation network (nodes: papers, links: one paper citing another),
	\item A$\leftrightarrow$A, directed author citation network (nodes: authors, links: one author cites another in at least one of his/her papers),
	\item A$-$A, undirected co-authorship network (nodes: authors, links: co-authorship of at least one paper).
\end{itemize}
This gives us the total of $6+6+6=18$ networks ($12$ directed and $6$ undirected), to which we devote the rest of this paper. Our goal is to study the consistency among the networks within each category in terms of their topologies, from which we draw conclusions on the consistency among the databases.

In Table\,\ref{table1} we summarize the basic properties of the $18$ examined networks. Numbers of nodes and links vary greatly, but are always larger than $10^4$. WCC is the fraction of nodes contained in the largest connected component (weak connectivity for directed networks, see Methods). With exception of DBLP P$\rightarrow$P network, it always contains at least $80\%$ of nodes (DBLP database consists mostly of the papers only from major journals and conferences, which rarely cite one another). Some papers/authors are never cited, others do not cite any other paper/author in the same database. Motivated by this, we consider ``bow-tie''~\cite{SFB14} of directed networks, which indicates the fraction of `core' nodes (both citing and cited), in contrast to the fraction of `in' nodes (never cited) and `out' nodes (not citing). Diversity of these parameters (note their independence from networks' sizes) already gives a hint at the variability among the databases. Some additional particularities: P$\rightarrow$P networks are in general acyclic since a paper can only cite older papers. Rare exceptions occur due to parallel publication of multiple papers citing one another, and due to errors. These networks include the information on chronology of publishing. In contrast, A$\leftrightarrow$A networks often contain cycles, since collaborating authors typically cite one another. Also, basically all nodes here will have self-loops (authors cite their previous work). On the other hand, no P$\rightarrow$P network node has a self-loop, since papers usually do not cite themselves (except in very unusual cases or due to errors).

\begin{table}[!hbt] \centering\small
        \caption{\textbf{Basic network measures.} The values of all basic network measures for the $18$ examined networks. See Methods for details on the definitions of network measures and their computation.}
	\begin{tabular}{llcccccc} \hline
		& & \multicolumn{2}{c}{Network size} & \multicolumn{4}{c}{Network bow-tie} \\\hline
		Type & Database & \# Nodes & \# Links & \% WCC & \% In & \% Core & \% Out \\\hline
		P$\rightarrow$P & APS & $\phantom{0,}450$,$084$ & $\phantom{0}4$,$691$,$938$ & $\phantom{0}99.8\%$ & $\phantom{0}2.6\%$ & $82.7\%$ & $14.5\%$ \\
		& WoS & $\phantom{0,}728$,$673$ & $\phantom{0}3$,$633$,$240$ & $\phantom{0}96.9\%$ & $11.5\%$ & $53.9\%$ & $31.5\%$ \\
		& DBLP & $1$,$467$,$987$ & $\phantom{0}1$,$502$,$092$ & $\phantom{00}4.3\%$ & $\phantom{0}0.6\%$ & $\phantom{0}0.6\%$ & $\phantom{0}3.1\%$ \\
		& PubMed & $5$,$853$,$635$ & $18$,$790$,$433$ & $\phantom{0}99.7\%$ & $89.9\%$ & $\phantom{0}4.3\%$ & $\phantom{0}5.5\%$ \\
		& Cora & $\phantom{0,}195$,$946$ & $\phantom{00,}608$,$475$ & $\phantom{0}99.0\%$ & $83.7\%$ & $\phantom{0}8.6\%$ & $\phantom{0}6.6\%$ \\
		& arXiv & $\phantom{0,}\phantom{0}27$,$770$ & $\phantom{00,}352$,$768$ & $\phantom{0}98.7\%$ & $\phantom{0}9.2\%$ & $73.6\%$ & $15.9\%$ \\\hline
		A$\leftrightarrow$A & APS & $\phantom{0,}260$,$816$ & $40$,$556$,$550$ & $100.0\%$ & $\phantom{0}1.7\%$ & $84.6\%$ & $13.7\%$ \\
		& WoS & $\phantom{0,}470$,$227$ & $20$,$291$,$830$ & $\phantom{0}99.5\%$ & $\phantom{0}9.9\%$ & $65.3\%$ & $24.4\%$ \\
		& DBLP & $\phantom{0,}\phantom{0}14$,$880$ & $\phantom{00,}219$,$173$ & $\phantom{0}98.8\%$ & $59.4\%$ & $26.8\%$ & $12.6\%$ \\
		& PubMed & $\phantom{0,}638$,$178$ & $11$,$905$,$813$ & $\phantom{0}99.8\%$ & $51.1\%$ & $31.2\%$ & $17.5\%$ \\
		& Cora & $\phantom{0,}\phantom{0}21$,$521$ & $\phantom{00,}582$,$021$ & $\phantom{0}99.6\%$ & $\phantom{0}9.2\%$ & $66.2\%$ & $24.1\%$ \\
		& arXiv & $\phantom{0,}\phantom{0}11$,$779$ & $\phantom{00,}586$,$562$ & $\phantom{0}99.4\%$ & $\phantom{0}7.4\%$ & $79.3\%$ & $12.7\%$ \\\hline
		A$-$A & APS & $\phantom{0,}248$,$866$ & $\phantom{0}4$,$231$,$131$ & $\phantom{0}90.0\%$ & - & - & - \\
		& WoS & $\phantom{0,}531$,$952$ & $\phantom{0}2$,$966$,$442$ & $\phantom{0}89.8\%$ & - & - & - \\
		& DBLP & $1$,$359$,$484$ & $\phantom{0}5$,$821$,$900$ & $\phantom{0}89.9\%$ & - & - & - \\
		& PubMed & $1$,$675$,$367$ & $16$,$926$,$075$ & $\phantom{0}96.4\%$ & - & - & - \\
		& Cora & $\phantom{0,}\phantom{0}23$,$480$ & $\phantom{00,}130$,$644$ & $\phantom{0}87.5\%$ & - & - & - \\
		& arXiv & $\phantom{0,}\phantom{0}11$,$868$ & $\phantom{00,}\phantom{0}24$,$638$ & $\phantom{0}81.4\%$ & - & - & - \\\hline
	\end{tabular} \captionsetup{font=small}
	\label{table1}
\end{table}

\begin{figure}[!hbt]
	\includegraphics[width=\textwidth]{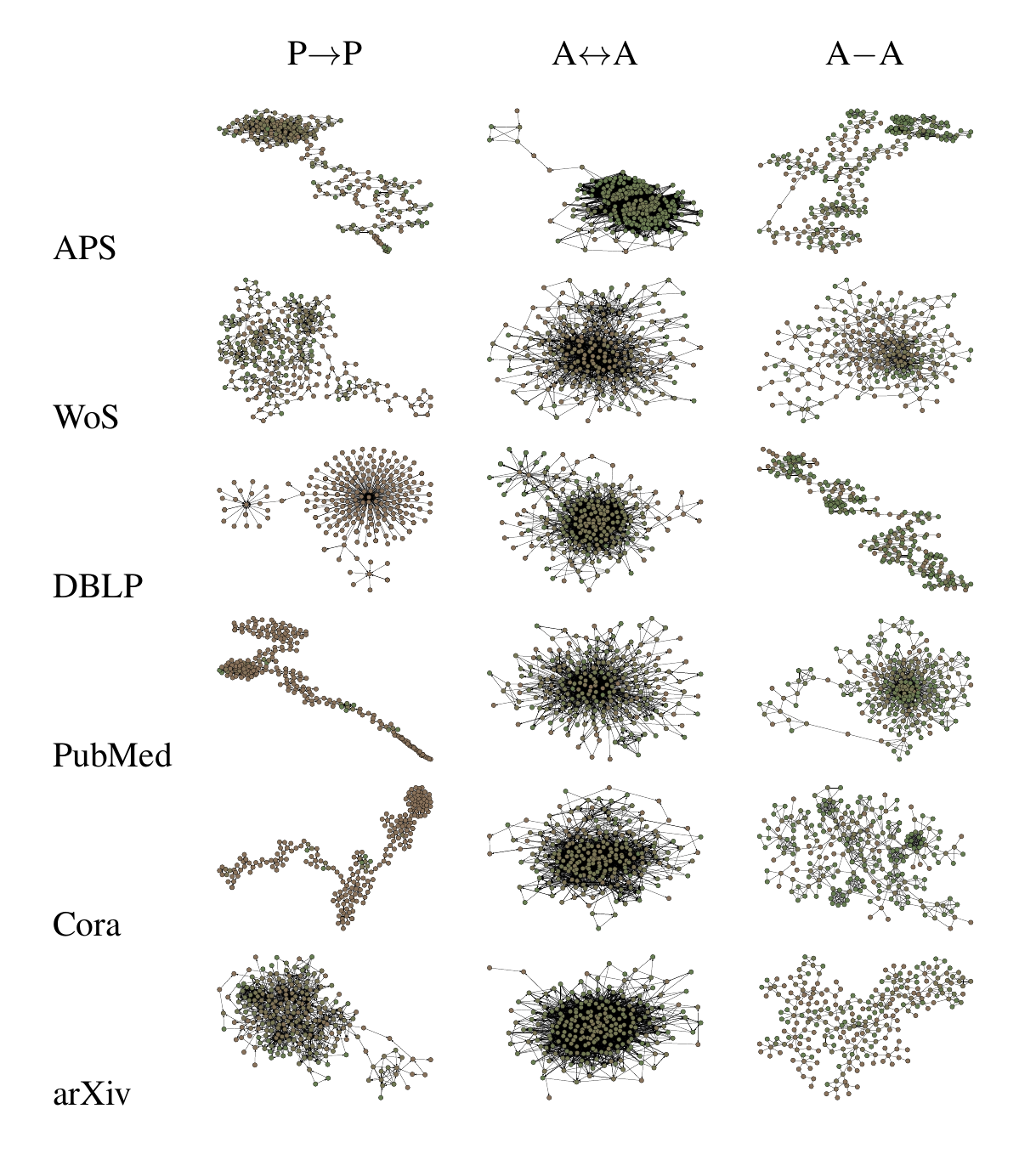}
	\caption{\textbf{Graphical visualization of the network samples.} As indicated, each sample corresponds to one of the $18$ examined networks. See Methods for details on network sampling algorithm.} \label{figure1}
\end{figure}

We now observe the following: while the three network paradigms (P$\rightarrow$P, A$\leftrightarrow$A and A$-$A) are all bibliometric in nature, the resulting network architectures are very different. In other words, by representing a database via three different network paradigms, we view its complexity from three different standpoints. These three representations are largely uncorrelated, each contributing some new information (for example, although collaborating authors often cite one another, they also cite other scientists they never worked with, and sometimes co-author papers with scientists they never cited or got cited by). This allows the comparison among the databases along three independent lines, allowing us to isolate for each database the network category best suited for its study. To illustrate this point, we graphically visualize a sample of each network in Fig.\,\ref{figure1}, obtained via network sampling algorithm~\cite{LF06,ANK11}. Network samples are small subnetworks which capture the key topological features of the corresponding large (complete) networks (visualizing complete networks is impractical due to their size, see Methods). Visual comparison of network samples coming from the same database (horizontal) indeed indicates that each network paradigm presents a database from a different angle, viewing its complexity from a specific aspect. Comparison of network samples corresponding to different databases (vertical) reveals significant topological differences among them. They exist along all three vertical columns, and are most clearly pronounced for P$\rightarrow$P and A$-$A networks. This suggests that in all three network categories there are at least some differences in the data structure and bibliometric precision among the databases. Motivated by this insight, we continue our study in more quantitative terms.

We begin by introducing a platform for quantification of the network topologies~\cite{SFB14}. On top of $6$ network measures introduced in Table\,\ref{table1}, we compute for each network additional $16$ measures:
\begin{itemize}
	\item Degree statistics and distribution parameters: $\left< k \right>$, $\gamma$, $\gamma_{in}$, $\gamma_{out}$,
	\item Degree mixing quantifiers: $r$, $r_{(in,in)}$, $r_{(in,out)}$, $r_{(out,in)}$, $r_{(out,out)}$,
	\item Clustering distribution parameters: $\left<c\right>$, $\left<b\right>$, $\left<d\right>$,
	\item Clustering mixing quantifiers: $r_{c}$, $r_{b}$, $r_{d}$,
	\item Effective diameter parameter: $\delta_{90}$.
\end{itemize}
The definition and interpretation of each network measure along with the procedure used for its computation are explained in Methods. The Supporting Information Fig.\,A in File S1 graphically shows relevant node degree and clustering profiles and distributions (see Methods). Rather than studying all the values (which are reported in the Supporting Information Tables B1 and B2 in File S1), we would here like to illustrate our approach to quantifying the mutual consistency of databases relying on these measures. We focus on a specific one among them, clustering mixing $r_{b}$, whose values for all networks are shown in Table\,\ref{table2}. Looking at the table row by row, three observations can be made. All P$\rightarrow$P networks are relatively consistent in their values except for DBLP. Similarly, with exception of APS, all A$\leftrightarrow$A networks are roughly consistent. Finally, PubMed is the only database not consistent with the others when it comes to A$-$A networks. This suggests a simple way to quantify the consistency of databases within each network category. Of course, we expect that the consistency will depend on the chosen network measure. Ideally, the ``best'' database would be the one most consistent with as many others for as many measures as possible. However, as we show in what follows, trying to identify such a database is elusive. Instead, our main result is the consistent quantification of their mutual consistency for each network category. Our findings are to be understood as an ``advice'' to researchers in bibliometrics about the suitability of various network paradigms in relation to the database of their interest.

%\rowcolors{1}{gray!1}{gray!20}
\begin{table}[!hbt] \centering\small
	\caption{\textbf{Values of clustering mixing.} Values of the network measure clustering mixing $r_{b}$ for all $18$ examined networks. See text for discussion.}
	\begin{tabular}{lcccccc} \hline
		& APS &  WoS     & DBLP    &  PubMed      & Cora      & arXiv \\ \hline
		P$\rightarrow$P & $0.43$ & $0.51$  &  $0.66$ &  $0.41$   & $0.43$  & $0.51$  \\
		A$\leftrightarrow$A & $0.71$ & $0.12$  &  $0.17$ &  $0.29$   & $0.34$  & $0.22$  \\
		A$-$A & $0.87$ & $0.91$  &  $0.84$ &  $0.46$   & $0.85$  & $0.64$  \\
		\hline
	\end{tabular} 	\captionsetup{font=small}
	\label{table2}
\end{table}

\begin{figure}[!hbt]
		\includegraphics[width=\textwidth]{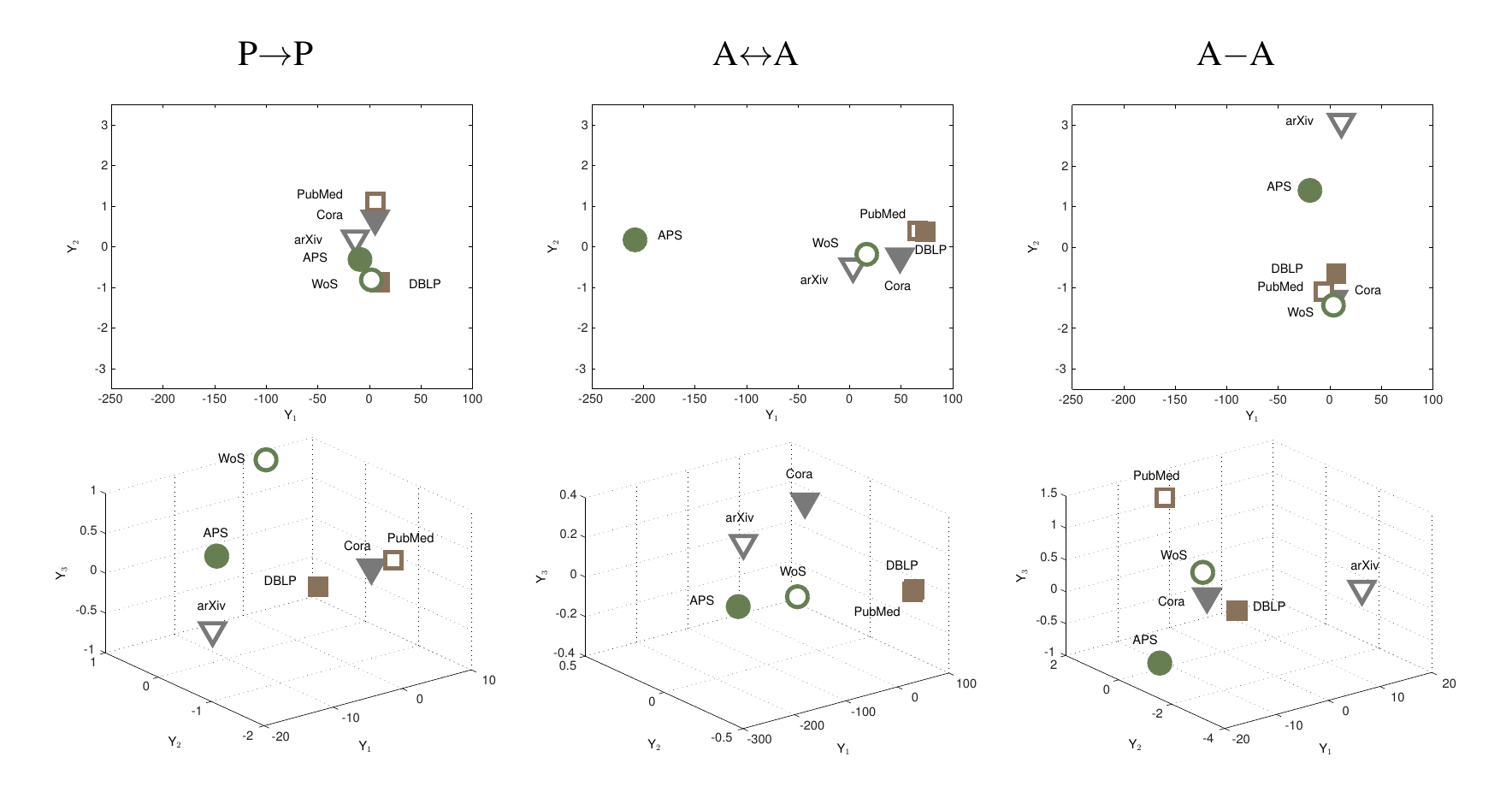}
	\caption{\textbf{Multidimensional scaling (MDS) analysis.} Embedding of points in 2D (top row) and 3D space (bottom row) obtained via MDS. Each point represents one database as indicated. Distance between any pair of points is representative of the average difference of network measure values for the corresponding database pair, and in adequate ratio with distances between other points in that plot.} \label{figure2}
\end{figure}

Our next step is to employ the standard technique of multidimensional scaling (MDS)~\cite{Cox2010,Martinez2004}, with aim to graphically visualize the overall differences among the databases. To this end, for each network category, we consider the differences of values of all network measures and for each pair of databases. The result of MDS is the embedding of $6$ points representing $6$ databases into the Euclidean space of given dimensionality. This embedding is done in a way that the Euclidean distance between each pair of points is representative of the inconsistency between the corresponding databases, in terms of the average difference in values of network measure (see Methods). The obtained embeddings for 2- and 3-dimensional space are shown in Fig.\,\ref{figure2}. Closer together databases are, better the overall consistency of their network measures. For the case of P$\rightarrow$P networks, only PubMed and Cora appear to be relatively consistent with one another. PubMed and DBLP display a nearly perfect consistency between them for A$\leftrightarrow$A networks, with some (independent) consistency among arXiv, WoS and Cora. For A$\leftrightarrow$A networks, best consistency is found for DBLP, Cora and WoS. Indeed, the consistency among databases is dependent on the network paradigm used to represent them. Even within each of these categories, it seems difficult to establish which databases are mutually consistent and which are not. In what follows, we seek to establish at least some approximate results in this direction.

\begin{figure}[!hbt]
		\includegraphics[width=\textwidth]{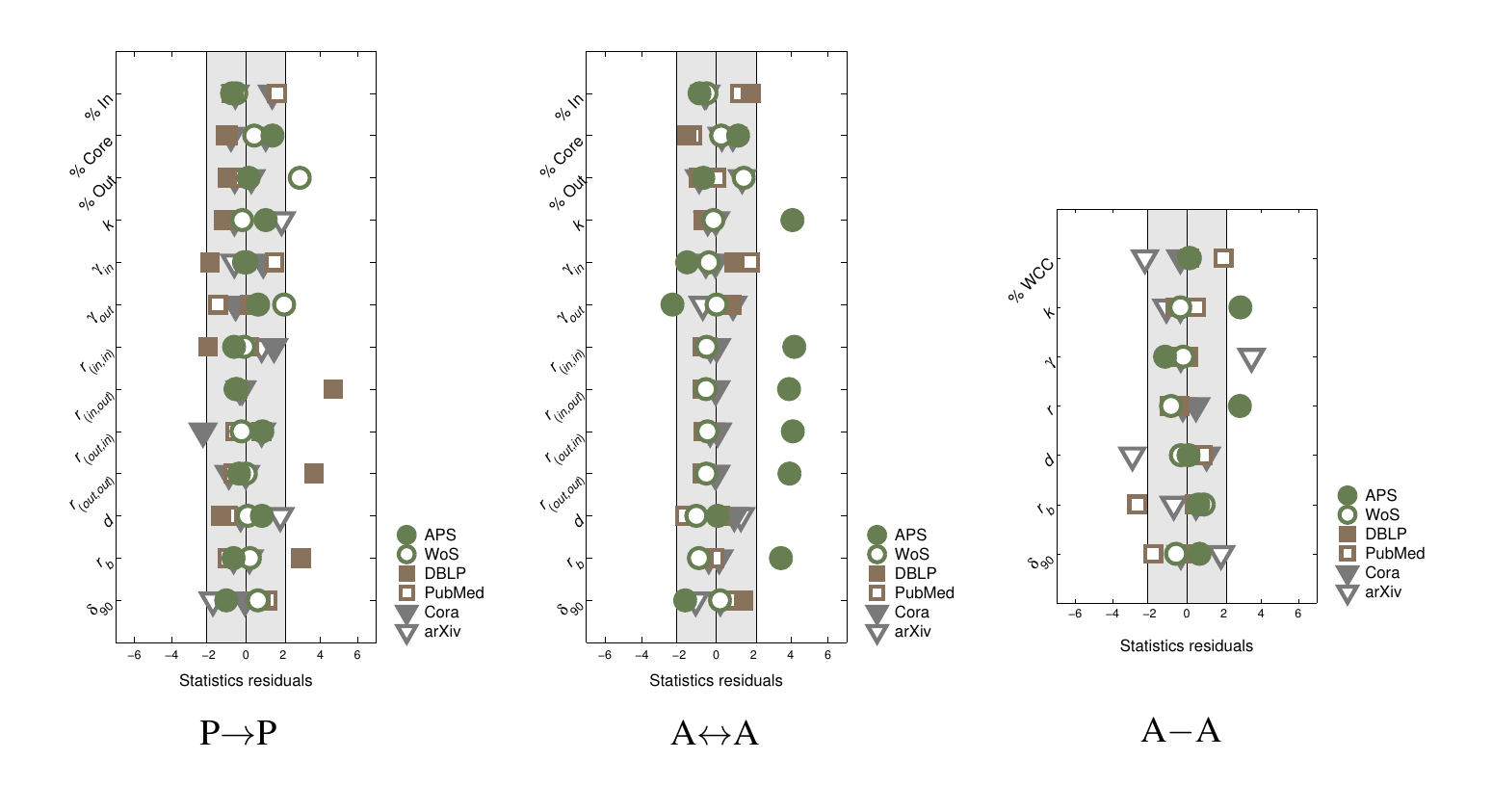}
	\caption{\textbf{Analysis via residual computation.} Externally studentized residuals for all databases, computed separately for each independent network measure and each network category. See Methods for interpretation and details on computation.} \label{figure3}
\end{figure}

Returning to the values of network measures, we construct another comparison among databases, this time relying on the standard statistical analysis. We begin by realizing that network measures are not all independent~\cite{WS98,SV05}, neither are the ``true'' values for any of them known. This calls for identifying a set of measures which cumulatively provide the optimal information on the network topologies. To this end, for each database we first compute the externally studentized residual, separately for each network measure and category (see Methods). We express the residuals in the units of standard deviations for that measure. That is to say, the database with residual zero is the one most ``in the middle'' according to that measure. Oppositely, the database with the residual farthest from zero is the one least surrounded by others. Next we use these residuals to identify the optimal set of independent network measures, separating between directed and undirected networks (Methods). We found this to consist of $13$ measures for directed and $7$ for undirected networks, whose residuals are reported in Fig.\,\ref{figure3}. We also confirmed that this selection still cumulatively provides enough information to enable the differentiation among the networks (Methods). The difference with the previous MDS analysis is that here we treat each network measure separately, without mixing their values in any way, and we also remove some measures as redundant. This is done not just to exclude possible inter-dependences among them, but also since the values belonging to different measures cannot always be directly compared. For P$\rightarrow$P networks, with exception of DBLP, all databases appear to be relatively consistent. A$\leftrightarrow$A networks also display good consistency, with exception of APS which shows a notable discrepancy. A$-$A networks reveal APS and arXiv databases to be most inconsistent with others. Note that these results are in a good agreement with the results of the MDS analysis (Fig.\,\ref{figure2}). In fact, the analysis of residuals again confirms that it is hard to identify a single ``best'' database in terms of biggest consistency with other databases, even within the realm of a single network category. Needless to say, it would be even more elusive to search for the ``best'' database simultaneously for all network categories.

\begin{figure}[!hbt]
		\includegraphics[width=\textwidth]{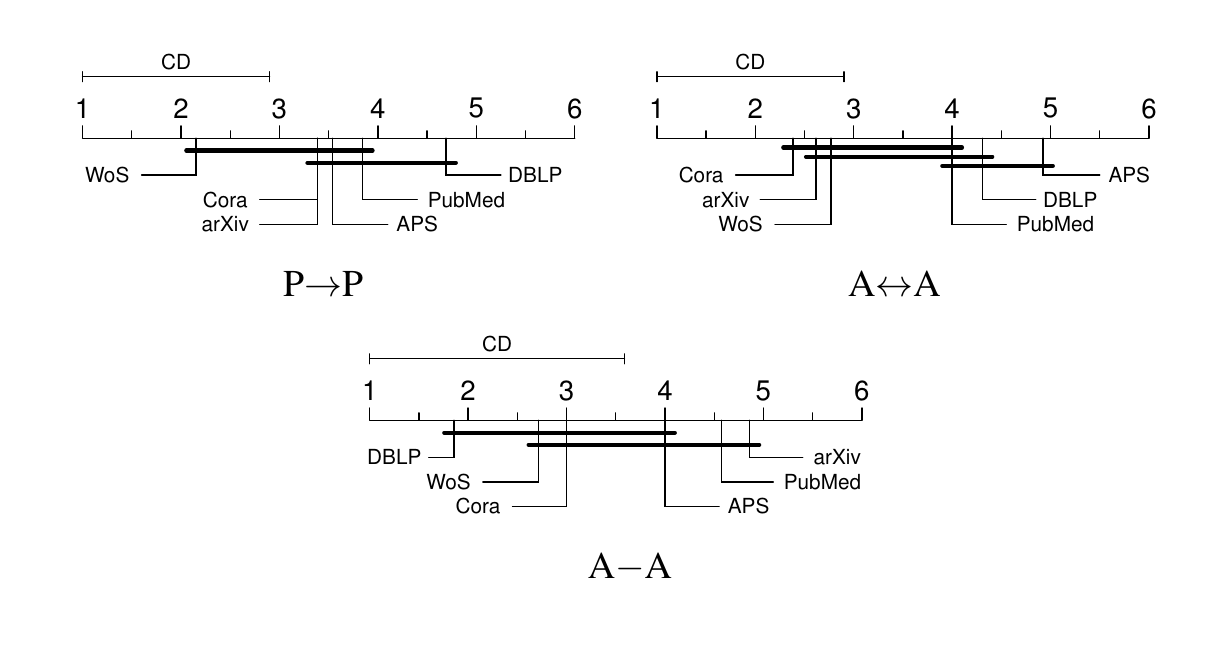}
	\caption{\textbf{Final overall ranking.} Ranking of databases for all three network categories. Critical difference (CD) indicates what range of ranking differences is not statistically significant. The difference in ranking of databases underlined by the common bold line are not statistically significant (Methods).} \label{figure4}
\end{figure}

Still, as our wish is to offer at least some qualitative argument on mutual consistency of databases, we construct the ranking of databases from computed residuals. Within each network category we proceed as follows. For each network measure, we assign the rank $1$ to the database with the residual closest to zero, rank $2$ to the database with the residual second closest to zero, and so on until we assign the rank $6$. Averaging these ranks yields an average rank for each database, defining a database ranking for each category (see Methods). Smaller the rank of a database, better its overall consistency with the rest. The rankings are reported in Fig.\,\ref{figure4}. However, despite a clear hierarchy given by ranking, not all ranking differences are statistically significant. To account for this, we indicate as CD (critical difference) the width corresponding to the $p$-value of $0.1$ by which we establish the statistical significance. Thus, any ranking difference smaller than CD is not statistically significant. For easier understanding of the figure, we add bold lines to indicate groups of databases where ranking differences are not statistically significant. For P$\rightarrow$P networks, WoS is the most consistent database, even though its ranking is not statistically different from Cora, arXiv, APS and PubMed. The same is visible from A$\leftrightarrow$A networks, where rankings of Cora and arXiv are even somewhat better than that of WoS. Finally, DBLP ranks best in terms of A$-$A networks, followed by WoS, Cora and APS, none of which are actually statistically worse. Based on the available data, these results represent the optimal differentiation among the databases in terms of their consistency. We believe that the differences we found are to be attributed to different methodologies in maintaining different databases. Specifically, WoS keeps track of citations manually, thus avoiding many errors related to referencing styles and authors' names, which to a large extent explains its good quality.

As mentioned earlier, bibliometric studies are in practice done by relying on the data that happens to be available to the researcher. These data usually comes from a single database, which is usually among here considered databases. However, such studies often suffer from criticism of bias coming from relying on a single database. To aid this situation, we reiterate the above results towards offering concrete suggestions regarding the choice of the network paradigm best suited for studying any given database. WoS can be basically studied via any network paradigm. Roughly the same can be said of Cora. When examining arXiv, one should avoid A$-$A networks. In contrast, study of DBLP should exactly go via A$-$A networks. On the other hand, studying APS and PubMed seems to be less promising. However, if the choice has to be made, P$\rightarrow$P appears to be the best option for both.

\section*{Discussion}

Our work was done relaying on the representative datasets from six databases which, to our best knowledge, are the ones most frequently used in modern bibliometrics. Of course, we realize that these by no means include all the relevant bibliometric data. In particular, some databases including SCOPUS, Google Scholar and CiteSeer are missing from our study. Unfortunately, we were unable to obtain the representative datasets from these databases. However, some of the missing databases rely on the bibliometric methodology similar to some of the studied databases (notably, SCOPUS uses methodology very similar to WoS~\cite{FPMP08,Fia11}). For this reason, we believe that the presence of these databases would not significantly alter our results. Furthermore, the considered databases do not always overlap in the scientific fields they cover (for example APS, Cora and PubMed). Due to this a minor bias could be present in our study, which unfortunately can never be entirely removed if one wants to compare different fields. On the other hand, all databases refer to computer and natural sciences, which are known to have very similar collaboration and citation cultures. We thus believe this bias had no major impact on our key findings. Nevertheless, we agree that there exists an intrinsic incomparability between distant scientific fields (for instance computer science and history), which necessitates new approaches and methodologies able to offer more objective comparisons. Another interesting question revolves around aggregation of the databases: aggregate data would provide a closer approximation of the ground truth, yet it might be hindered by the above described discrepancies in the datasets. We leave this open problem for future work.

One could argue that bibliometric networks are not the only framework for studying the consistency among scientific databases. For example, a simple comparison of a sample of records could provide insights on their precision. Yet, complex networks have become over the years a well established platform for investigating complex systems. This is due to their power to reveal the information hidden in the shear complexity of systems such as scientific community. For this reason, while acknowledging the value of additional approaches to this problem, we argue that networks are presently the most appropriate framework. On the other hand, our study could be extended to other network paradigms used for bibliometric networks, such as those based on linking the papers which share keywords or specific words in the title or abstract~\cite{kastrin2014}.

The main ingredient of our methodology is the network comparison, realized via computation of $22$ network measures and identifying the independent among them. In fact, this turns out to be the simplest approach, easily applicable to both directed and undirected networks. However, we note that the NP-hard problem of network comparison is a topic of constant interest in the field, with novel ideas rapidly accumulating~\cite{natasa2014}. Also, our approach was largely based on classical statistical analysis involving significance testing, which was recently scrutinized~\cite{Ziliak2008}. However, besides being in agreement with our previous paper~\cite{SFB14}, our results are also confirmed by MDS analysis which is in no way related to classical statistics. We thus argue that our statistical results are indeed informative. Finally, while noting that improvements of our methodology are possible, we hope our work traces a new avenue for all interested in critically examining science as a human endeavor.

\section*{Methods}
\label{Methods}

\paragraph{The data.} The data has been extracted from publicly available repositories and purchased from commercial bibliographic sources. Authors and publications neither citing nor cited were discarded, together with authors not collaborating. Self-citations of papers that occur due to errors were discarded. The details on six studied databases are below. \\

\noindent \textit{American Physical Society} (APS) is the world's second largest organization of physicists ({\it http://www.aps.org}), behind German DPG. It publishes a range of scientific journals, including the Physical Review series, Physical Review Letters and Reviews of Modern Physics. The data considered here contains all publications in aforementioned journals up until 2010 consisting of $450$,$084$ papers and $264$,$844$ authors, and $4$,$710$,$547$ citations between the papers. \\

\noindent \textit{Web of Science} (WoS) is informally considered the most accurate scientific bibliographic database, professionally hand-maintained by Thomson Reuters ({\it http://thomsonreuters.com}). It dates back to early $1950$s~\cite{Gar55,Pri65} and contains over $45$ million~records of publications from all fields of science~\cite{Fia11}. For this study, we consider all publications in WoS category Computer Science up until late 2014. The entire dataset includes $978$,$821$ papers and $580$,$112$ authors, and $3$,$633$,$421$ citations between the papers. \\

\noindent \textit{DBLP Computer Science Bibliography} (DBLP) indexes major journals and proceedings from all fields of computer science~\cite{Ley02} ({\it http://dblp.uni-trier.de}). It is freely available since $1993$ and hand-maintained by University of Trier, Germany. It contains more than $2$ million records of publications, while the citation information is rather scarce compared to WoS~\cite{Fia11}. For this study, we considered a snapshot of the database on September 2014 including $2$,$696$,$491$ papers and $1$,$424$,$895$ authors, and $1$,$534$,$369$ citations between the papers ({\it http://lovro.lpt.fri.uni-lj.si}). \\

\noindent \textit{PubMed} (PubMed) is a search engine of MEDLINE database focusing on life sciences and biomedicine, maintained by US National Institutes of Health ({\it http://www.ncbi.nlm.nih.gov}). It contains about $24$ million citations between publications dating back to late 19th century. For this study, we extracted open access publications from PubMed Central Collection up until 2014 and author information from MEDLINE Baseline Repository between 2012 and 2014. We thus obtained $5$,$853$,$635$ papers and $1$,$716$,$762$ authors, and $18$,$842$,$120$ citations between the papers. \\

\noindent \textit{Computer Science Research Paper Search Engine} (Cora) is a service for automatic retrieval of publication manuscripts from the Web using machine learning techniques~\cite{MNRS00}. It contains over $200$,$000$ publication records collected from the websites of computer science departments at major universities in August $1998$ ({\it http://people.cs.umass.edu/ $\tilde{ }$mccallum}). For this study, we consider a complete database including $195$,$950$ papers and $24$,$911$ authors, and $623$,$287$ citations between the papers ({\it http://lovro.lpt.fri.uni-lj.si}). \\

\noindent \textit{arXiv.org} (arXiv) is a public preprint repository of publication drafts uploaded by the authors prior to an actual journal or conference submission hosted by the Cornell University in US since $1991$~\cite{Gin11} ({\it http://arxiv.org}). It currently contains almost one million publications from physics, mathematics, computer science and other fields. For this study, we consider all publications in arXiv category High Energy Physics Theory between 1992 and 2003 ({\it http://snap. stanford.edu}). The data contains $27$,$770$ papers and $12$,$820$ authors, and $352$,$807$ citations between the papers.

\paragraph{Network sampling algorithm.} The goal of network sampling is to extract a subnetwork from the complete (often very large) network, which would be representative of its topological (or other) properties. Due to its small and regulable size, this subnetwork (which we call network sample) can be easily visualized and compared to network samples representing other networks. We obtained the network samples by considering the induced subgraphs on the nodes visited by a random walker starting at some random node~\cite{LF06,ANK11}. That is to say, our network sample includes all the nodes visited by the walker after some number of steps, together with all the links connecting those nodes. In fact, this has been proven to generate samples that are most similar to the original networks~\cite{BSB14}. In our work we generated $5000$ networks samples of $250$ nodes for each of the original networks, whereas the best sample is selected according to Kolmogorov-Smirnov distance between the degree distributions.

\paragraph{The network measures.} To quantify the topology of the examined networks we used 22 different measures. Below we explain the remaining 20 measures (number of nodes and links is obvious). For undirected networks we compute only the measures naturally defined for them. For directed networks, upon computing the measures naturally defined for them, we disregard their directionality, and also compute the measures normally referring to undirected networks. Largest (weakly) connected component of a directed network is its maximal subnetwork such that all its nodes are mutually reachable, disregarding the directionality. We define as WCC the size of this subnetwork. We measured the strong connectivity only in the context of network bow-tie structure~\cite{SFB14} (\% core, \% in, and \% out). \\

\noindent \textit{Degree distributions.} For directed networks, in-degree $k_{in}$ and out-degree $k_{out}$ of a node are respectively the number of incoming and outgoing links. $k$ is the degree of a node, $k=k_{in}+k_{out}$, and $\left<k\right>$ denotes the mean degree. For undirected networks we deal only with $k$. We computed the exponents $\gamma_{in}$, $\gamma_{out}$ and $\gamma$ which characterize the degree distributions (for directed network $\gamma$ is computed disregarding the directionality). This is done by fitting the tails of the distributions by maximum-likelihood estimation: 
\begin{equation}
	\gamma_{\cdot}=1+n\left(\sum_{V}\ln\left.k_{\cdot}/k_{min}\right.\right)^{-1} \;\;\; \mbox{for} \;\;\; k_{min}\in\{10,25\}. 
\end{equation}
In cases exhibiting power-law degree distributions, these exponents correspond to the actual power-law exponents. In all cases these exponents were characteristic of the degree distributions, in the sense that similar distributions have similar exponents. \\

\noindent \textit{Degree mixing.} Neighbor connectivity $N{k_{\cdot}}$ is the mean neighbor degree of all network nodes with degree $k_{\cdot}$~\cite{PVV01}. The degree mixing $r_{(\alpha,\beta)}$ is the Pearson correlation coefficient of $\alpha$-degrees or $\beta$-degrees at links' source and~target nodes,~respectively~\cite{FFGP10}:
\begin{equation}
	r_{(\alpha,\beta)} = \frac{1}{\sigma_{k_{\alpha}}\sigma_{k_{\beta}}}\sum_{L}\left(k_{\alpha}-\left<k_{\alpha}\right>\right)\left(k_{\beta}-\left<k_{\beta}\right>\right), 
\end{equation}
where $\left<k_{\cdot}\right>$ and $\sigma_{k_{\cdot}}$ are the means and standard deviations, $\alpha,\beta\in\{in,out\}$ (measured only for directed networks). $r$ is the mixing of degrees $k$, measured for undirected networks and for directed ones disregarding their directionality~\cite{New03a}. \\

\noindent \textit{Clustering distributions and mixing.} All clustering coefficients were computed disregarding the directionality of directed networks. Clustering coefficient $c$ is usually defined as the link density of its neighborhood~\cite{WS98}:
\begin{equation}
	c = \frac{2t}{k(k-1)}, 
\end{equation}
where $t$ is the number of linked neighbors and $k(k-1)/2$ is the maximum possible number, $c=0$ for $k\leq 1$. The mean $\left<c\right>$ is denoted network clustering coefficient~\cite{WS98}, while the clustering mixing $r_{c}$ is defined as before. Clustering profile gives the mean clustering $C{k}$ of nodes with degree $k$~\cite{RSMOB02}. Note that the denominator in the equation above introduces biases when $r<0$~\cite{SV05}. Thus, we rely on delta-corrected clustering coefficient~$b$, defined as $c\cdot k/\Delta$~\cite{NMB05}, where $\Delta$ is the maximal degree $k$ and $b=0$ for $k\leq 1$. Similarly, degree-corrected clustering coefficient $d$ is defined as $t/\omega$~\cite{SV05}, where $\omega$ is the maximum number of linked neighbors with respect to their degrees $k$ and $d=0$ for $k\leq 1$. From definition it follows  $b \leq c \leq d$. \\

\noindent \textit{Diameter statistics.} All diameter statistics were computed disregarding the directionality of directed networks. Hop plot shows the percentage of mutually reachable pairs of nodes $H(\delta)$ with $\delta$ hops~\cite{LKF07}. The network diameter is defined as the minimal number of hops $\delta$ for which $H(\delta)=1$, while the effective diameter $\delta_{90}$ is the number of hops at which $90\%$ of such pairs of nodes are reachable~\cite{LKF07}, $H(\delta_{90})=0.9$. Hop plots are averaged over $100$ realizations of the approximate neighborhood function with $32$ trials~\cite{PGF02}.

\paragraph{Multidimensional scaling (MDS).} MDS is a statistical technique that visualizes the level of similarity of individual objects of a dataset. From the range of the available MDS techniques, we used the non-metric multidimensional scaling (NMDS), which work as follows. Given are $h$ objects (or points) defined via their coordinates in $l$ dimensions. This situation is expressed via $h \times l$ matrix called $H$. From this original matrix $H$ we compute the dissimilarity $h \times h$ matrix $D$, in which each matrix element $D(i,j)$ represents the Euclidean distance between the pair of objects $i$ and $j$ in the original matrix $H$. NMDS reduces the dimensionality of the problem, by transforming the $h \times h$ matrix $D$ into a $h \times p$ matrix $Y$, where $h$ is the number of objects (or points), now embedded in $p$ dimensions instead of $l$ ($p < l$)~\cite{Martinez2004}. The Euclidean distances between the obtained $h$ points in $Y$ are a monotonic transformation of the points in $D$ in $p$ dimensions. In our analysis, we used a original matrix $H$ with size of $6 \times 20$, meaning that the number of points (data basis) is $h=6$ and the number of coordinates is $l=20$. The original matrix $H$ is transformed into dissimilarity matrix $D$ with size of $6 \times 6$. Using NMDS we transformed the matrix $D$ into two matrices $Y'$ and $Y''$, so that $Y'$ has a size of $6 \times 2$, and $Y''$ has a size of $6 \times 3$.

\paragraph{Externally studentized residuals.}  Let $x_{ij}$ be the value of $j$-th network measure of $i$-th database, where $N$ is the number of databases, $N=6$. Corresponding externally studentized residual $\hat{x}_{ij}$~is:
\begin{equation}\label{eq:x}
	\hat{x}_{ij}=\frac{x_{ij}-\hat{\mu}_{ij}}{\hat{\sigma}_{ij}\sqrt{1-1/N}},
\end{equation}
where $\hat{\mu}_{ij}$ and $\hat{\sigma}_{ij}$ are the sample mean and the corrected standard deviation excluding the considered $i$-th database, $\hat{\mu}_{ij}=\sum_{k\neq i} x_{kj}/(N-1)$ and $\hat{\sigma}_{ij}^2=\sum_{k\neq i}(x_{kj}-\hat{\mu}_{ij})^2/(N-2)$. Assuming that the errors in $x$ are~independent and normally distributed, the residuals $\hat{x}$ have Student $t$-distribution with $N-2$ degrees of freedom. Significant differences in individual statistics $x$ are revealed by the independent two-tailed Student $t$-tests~\cite{CW82} at $P\mbox{-value}=0.1$, rejecting the null hypothesis $H_0$ that $x$ are consistent across the databases, $H_0:\hat{x}=0$. Thus, $\hat{x}_{ij}$ express the consistency of the database $i$ with the other databases, along the $j$-th network measure. Note also that the absolute values of individual residuals $|\hat{x}|$ imply a ranking $R$ over the databases, where the database with the lowest $|\hat{x}|$ has rank one, the second one has rank two and the one with the largest $|\hat{x}|$ has rank $N$.

\paragraph{Identifying independent network measures.}  Denote $r_{ij}$ to be the Pearson product-moment correlation coefficient of the residuals $\hat{x}$ for $i$-th and $j$-th network measure over all databases. Spearman rank correlation coefficient $\rho_{ij}$ is defined as the Pearson coefficient of the ranks $R$ for $i$-th and $j$-th statistics. Under the null hypothesis of statistical independence of $i$-th and $j$-th statistics, $H_0:\rho_{ij}=0$, adjusted Fisher~transformation~\cite{Fis15}:
\begin{equation}\label{eq:z}
	\frac{\sqrt{N-3}}{2}\ln\left.\frac{1+r_{ij}}{1-r_{ij}}\right.
\end{equation}
approximately follows a standard normal distribution. Pairwise independence of the selected network measures is thus confirmed by the independent two-tailed $z$-tests. %at $\pval=0.01$. 
This gives 13 independent measures for directed, and 7 independent measures for undirected networks, as shown in the Fig.\,\ref{figure3}. Furthermore, Friedman rank test~\cite{Fri37} confirms that chosen set of measures exhibits significant internal differences, as to still be informative on the databases (see below).

\paragraph{Ranking of databases.}  Significant inconsistencies between the databases are exposed using the methodology introduced for comparing classification algorithms over multiple data sets~\cite{Dem06}. Denote $R_i$ to be the mean rank of $i$-th database over the selected measure, $R_i=\sum_j R_{ij} / K$, where $K$ is the number of independent measures $K\in\{7,13\}$. One-tailed Friedman rank test~\cite{Fri37,Fri40} first verifies the null hypothesis that the databases are statistically equivalent and thus their ranks $R_i$ should equal, $H_0:R_i=R_j$. Under the assumption that the selected statistics are indeed independent, the Friedman testing statistic~\cite{Fri37}:
\begin{equation}\label{eq:F}
	\frac{12 K}{N(N+1)}\left(\sum_i R_i^2-\frac{N(N+1)^2}{4}\right)
\end{equation}
has $\chi^2$-distribution with $N-1$ degrees of freedom. By rejecting the hypothesis at $P\mbox{-value}=0.1$, we proceed with the Nemenyi post-hoc test that reveals databases whose ranks $R_i$ differ more than the critical~difference~\cite{Nem63}:
\begin{equation}\label{eq:cd}
	q\sqrt{\frac{N(N+1)}{6 K}},
\end{equation}
where $q$ is the critical value based on the studentized range statistic~\cite{Dem06}, $q=2.59$ at $P\mbox{-value}=0.1$. A critical difference diagram plots the databases with no statistically significant inconsistencies in the selected statistics~\cite{Dem06}.

\section*{Supporting Information}

% Include only the SI item label in the subsection heading. Use the \nameref{label} command to cite SI items in the text.

\subsection*{S1 File}
\textbf{Degree and clustering graphical profiles, continuation of network measures.} Node degree and clustering profiles and distributions of all the considered networks, along with other network statistics. See Methods for interpretation and details on computation.

\section*{Acknowledgments}
Authors thank American Physical Society and Thomson Reuters for providing the data. Thomson Reuters had no role in study design, data collection and analysis, decision to publish, or preparation of the manuscript. Work supported by Creative Core FISNM-3330-13-500033 funded by the European Union and The European Regional Development Fund, by the H2020-MSCA-ITN-2015 project COSMOS 642563, by the Slovenian Research Agency (ARRS) via programs P2-0359, P1-0383, via projects J1-5454, L7-4119, and by the Slovenian Ministry of Education, Science and Sport grant 430-168/2013/91. We thank the colleagues Dalibor Fiala, Ludo Waltman and Nees Jan van Eck for useful comments and discussions.

%\nolinenumbers

\includepdf[pages=-]{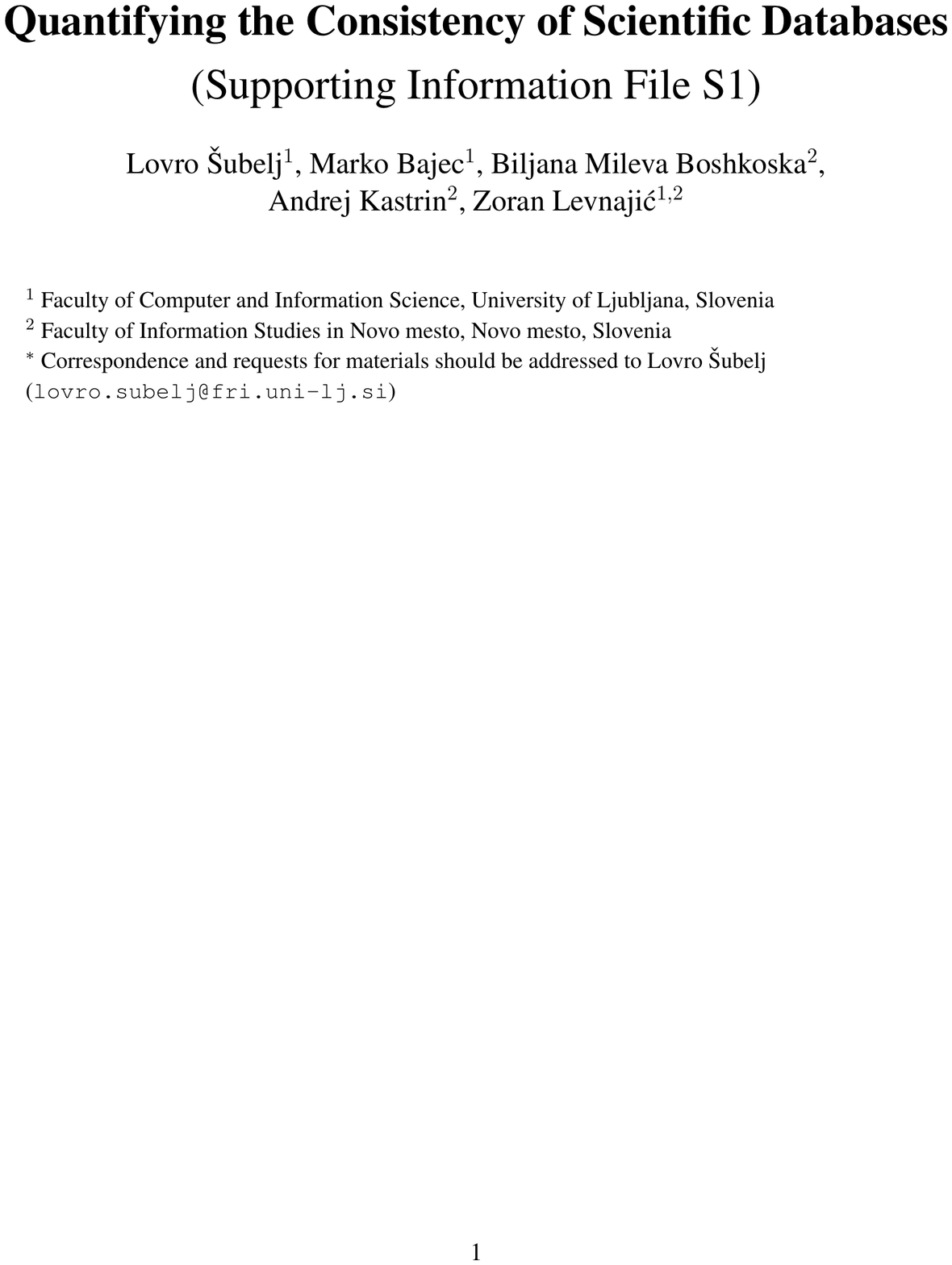}

\end{document}